\documentclass[12pt,a4]{article}
\usepackage{amssymb}
\usepackage{amsmath}
\newcommand{\wt}{\widetilde}
\newcommand{\ol}{\overline}
\newcommand{\Spin}{{\rm Spin}}
\newcommand{\U}{{\rm U}}
\newcommand{\SU}{{\rm SU}}
\newcommand{\SO}{{\rm SO}}

\begin{document}

\begin{titlepage}
    \begin{normalsize}
     \begin{flushright}
                 UT-04-16\\
                 {\tt hep-th/0406070}\\
                 June 2004
     \end{flushright}
    \end{normalsize}
    \begin{Large}
       \vspace{0.6cm}
       \begin{center}
       \bf Ricci-flat deformation of orbifolds and
       localized tachyonic modes
       \end{center}
    \end{Large}

\begin{center}
{
           Yosuke Imamura,
           \footnote{E-mail :
              imamura@hep-th.phys.s.u-tokyo.ac.jp}
           Fumikazu Koyama,
           \footnote{E-mail :
              koyama@hep-th.phys.s.u-tokyo.ac.jp}
and           Ryuji Nobuyama
           \footnote{E-mail :
              nobuyama@hep-th.phys.s.u-tokyo.ac.jp}
}
      \vspace{12mm}
{\large

              {\it Department of Physics,}\\[0.7ex]
              {\it University of Tokyo,}\\[0.7ex]
              {\it Hongo, Tokyo 113-0033, Japan}
}
      \vspace{1.2cm}
\end{center}
\begin{abstract}
We study Ricci-flat deformations of orbifolds in type II theory.
We obtain a simple formula for mass corrections to the twisted modes 
due to the deformations, 
and apply it to originally tachyonic and massless states in several examples.
In the case of supersymmetric orbifolds, 
we find that 
tachyonic states appear when the deformation breaks all the supersymmetries.
We also study nonsupersymmetric orbifolds 
${\bf C}^2/{\bf Z}_{2N(2N+1)}$, which is T-dual to $N$
type 0 NS5-branes. 
For $N\ge 2$, we compute mass corrections for states, 
which have string scale tachyonic masses. 
We find that the corrected masses coincide to ones 
obtained by solving the wave equation for the tachyon field 
in the smeared type 0 NS5-brane background geometry.
For $N=1$, we show that the unstable mode representing the bubble creation 
is the unique tachyonic mode. 
\end{abstract}

\end{titlepage}

\section{Introduction}
Tachyon condensation is one of main topics in string theory
actively investigated in the recent years.
The connection between open string tachyons
and the instability of D-branes conjectured by Sen \cite{Sen1,Sen2,Sen3,Sen4}
has made great progress in our understanding of dynamics of D-branes.

On the other hand we know much less about the closed string tachyon condensation.
One of the reasons is the close connection between closed strings and the spacetime.
We would expect some drastic change of the background spacetime itself
in the process of condensation.
Therefore it seems difficult
to analyze the closed tachyon condensation using conventional ways.
However, if closed tachyons exist only as localized states,
we can discuss more or less the fate of the spacetime when
tachyons condensate.

Several types of closed tachyons have been discussed.
In \cite{Melvin,fluxbrane}, relation between instability of fluxbranes
and closed string tachyon condensation is discussed.
In \cite{panic}, it is suggested that
condensation of tachyonic twisted modes
localized at fixed points of non-supersymmetric orbifolds
like ${\bf C}/{\bf Z}_n$ and ${\bf C}^2/{\bf Z}_n$
give rise to deformations of geometry of the background
spacetimes.
Configurations bringing with localized tachyons are also discussed in
many works
\cite{tseytlin}-\cite{mnr}.
Recently, the potentials of localized tachyons has been computed
by string field theory \cite{oz}.

In general, tachyonic twisted modes at an orbifold fixed point
have tachyonic masses of
the order of the string mass scale $\alpha'^{-1/2}$.
This implies that
we cannot directly analyze the decay process itself associated with
these tachyons without the help of string theory.
If we have localized tachyons with milder tachyonic masses
which does not include the string mass scale,
it may be possible to treat the condensation process in some classical way.
To obtain such modes,
let us consider an orbifold with massless twisted modes,
and perturb the background geometry so that it has non-vanishing
curvature of the order of $1/L^2$ with $L$ much larger than $\alpha'^{1/2}$.
Then we expect the twisted modes, which were originally massless,
receive mass corrections $\delta M^2\sim 1/L^2$,
and it can be negative depending on the way of
deformation.
If $L$ is sufficiently large,
it is expected for us to be able to analyze this instability
with classical (super)gravity.

In this paper
we demonstrate that we can actually obtain
twisted states with small tachyonic masses
in this way.
We consider several abelian orbifolds ${\bf R}^{10-d}\times{\bf R}^d/{\bf Z}_n$
with $d=4,6$ and $8$, and discuss the deformation of the metric of
the ${\bf R}^d/{\bf Z}_n$ factor.
Other fields (dilaton, $B$-field, metric of ${\bf R}^{10-d}$, etc.)
will be kept intact.
The equation of motion for the metric demands that the
deformed background is Ricci-flat.
This also guarantees the conformal symmetry on worldsheets.
We compute the mass correction to twisted modes
with the help of worldsheet non-linear $\sigma$-model.

In \cite{unstablens5}, the T-duality between
type 0 NS5-branes and an orbifold deformed
in this way is discussed,
and
the existence of
a tachyonic mode responsible to the creation of
Witten's bubble are shown.
However, it is difficult to find all the
unstable modes and to determine their tachyonic masses
by analyzing small fluctuations of the metric.
Indeed only the existence of at least one tachyonic mode was shown
in \cite{unstablens5}, and the number of all localized tachyons and their
masses including numerical factors were not determined.
One purpose of this paper is to give an answer to this problem
with the help of string theory.
Although the small tachyonic mass enables us to use classical gravity,
and the geometric interpretation is clarified,
string theory is still a powerful tool to compute the
masses of the tachyonic modes because
the correspondence between massless twisted modes
of supersymmetric orbifolds and
geometric deformations of the geometry has been well understood \cite{dm}.

This paper is organized as follows.
In the next section, we obtain a general formula for the
mass correction.
We apply the formula to several examples of
supersymmetric orbifolds in section
\ref{susy.sec}.
We see that we can obtain tachyonic modes when the deformation breaks
all the supersymmetry.
In section \ref{nonsusy.sec}, we discuss Taub-NUT deformation
of non-supersymmetric orbifolds, which is denoted by
${\bf C}^2/{\bf Z}_{2N(2N+1)}$ with the notation in \cite{panic},
and their relations to type 0 NS5-branes.
The last section is devoted to conclusions and discussions.

\section{Ricci-flat deformation}
Type II strings on curved backgrounds are described by
the supersymmetric non-linear $\sigma$-model
\begin{eqnarray}
L
&=&-\frac{1}{4\pi\alpha'}\int_0^{2\pi} d\sigma
\Big[ G_{mn}(\Phi)(\partial_\mu\Phi^m\partial^\mu\Phi^n
+i\Psi^m\gamma^\mu D_\mu\Psi^n)
\nonumber\\&&
\hspace{7em}+\frac{1}{6}R_{mpnq}(\Phi)(\Psi^m\Psi^n)(\Psi^p\Psi^q)
\Big],
\label{nlsm}
\end{eqnarray}
where $\Phi^m$ are scalar fields representing the coordinates
in the target space, and their superpartners $\Psi^m$ are
two-component Majorana spinors.
The metric $G_{mn}(\Phi)$ and the curvature tensor $R_{mnpq}(\Phi)$ depend on
the coordinates $\Phi^m$.
In order not to spoil the conformal invariance,
the background must be Ricci-flat.
Because of the Ricci-flatness of the background spacetime,
the curvature tensor $R_{mnpq}$ is identical to the Weyl tensor.
It is known that the number of independent components of the
Weyl tensor in $d$-dimension ($d\geq3$) is $d(d+1)(d+2)(d-3)/12$.

We treat the background curvature as perturbations to
flat orbifolds,
and our aim in this section is to give a general formula
for the mass corrections of twisted modes
up to first order in the curvature.
First task we have to do is to
expand the action (\ref{nlsm}) as $L=L_{(0)}+L_{(1)}+\cdots$,
where $L_{(0)}$ and $L_{(1)}$ are the free part and
the leading interaction terms proportional to the
curvature tensor, respectively.
The ellipsis represent higher order terms, which we are not interested in here.
The free equations of motion for $\Phi^m$ and $\Psi^m$ obtained from
$L_{(0)}$ are solved by
\begin{equation}
\Phi^m(\sigma^+,\sigma^-)=\phi^m(\sigma^+)+\wt\phi^m(\sigma^-),\quad
\Psi^m(\sigma^+,\sigma^-)=\left(\begin{array}{c}\psi^m(\sigma^+)\\ \wt\psi^m(\sigma^-)\end{array}\right).
\label{freesol}
\end{equation}
We consider twisted modes with twisted boundary conditions
\begin{equation}
\Phi^m(\sigma+2\pi)=U\Phi^m(\sigma)U^{-1},\quad
\Psi^m(\sigma+2\pi)=U\Psi^m(\sigma)U^{-1},\label{upu}
\end{equation}
where $U$ is an element of the orbifold group.
The OPEs for the left moving parts $\phi^m$ and $\psi^m$ are
\begin{equation}
\phi^m(\sigma)\partial_+\phi^n(0)
=\frac{\alpha'G^{mn}}{2\sigma}+\mbox{regular},\quad
\psi^m(\sigma)\psi^n(0)
=\frac{-i\alpha'G^{mn}}{\sigma}+\mbox{regular}.
\label{opes}
\end{equation}
We also have similar OPEs for $\wt\phi^m$ and $\wt\psi^m$.
From these OPEs and the boundary conditions (\ref{upu}),
we can determine the propagators including the regular terms
unambiguously.

The interaction term $H_{\rm int}=-L_{(1)}$ in the Hamiltonian is
obtained by the normal coordinate expansion of the action (\ref{nlsm}).
\begin{eqnarray}
H_{\rm int}=-L_{(1)}
&=&\int_0^{2\pi}\frac{d\sigma}{2\pi\alpha'}\Big[-\frac{1}{6}R_{mpnq}\Phi^p\Phi^q\partial_\mu\Phi^m\partial^\mu\Phi^n
\nonumber\\
&&-\frac{i}{6}R_{mpnq}\Phi^p\Phi^q(\Psi^m\gamma^\mu\partial_\mu\Psi^n)
+\frac{i}{4}R_{pqmn}(\Phi^p\partial_\mu\Phi^q)(\Psi^m\gamma^\mu\Psi^n)
\nonumber\\
&&+\frac{1}{12}R_{mpnq}(\Psi^m\Psi^n)(\Psi^p\Psi^q)\Big].
\label{Lint3}
\end{eqnarray}
$R_{mnpq}$ in (\ref{Lint3}) denotes the curvature tensor at
the orbifold fixed point $\Phi^m=0$.
Note that the operator products in (\ref{Lint3})
are not singular at the contact limit.
As we see in (\ref{opes}), the singular parts of OPEs
are proportional to the inverse metric $G^{mn}$.
Because of the Ricci-flatness of the background,
this singularity does not contribute to $H_{\rm int}$.

Let $J^{mn}$ denote anti-hermitian currents
associated with the targetspace rotational symmetry,
$K^{mn}$ and $\wt K^{mn}$ its left and right moving parts, respectively.
They are given by
\begin{equation}
K^{mn}
=\frac{1}{\alpha'}(2i\phi^{[m}\partial_+\phi^{n]}
   +\psi^{[m}\psi^{n]}),\quad
\wt K^{mn}
=\frac{1}{\alpha'}(2i\wt\phi^{[m}\partial_-\wt\phi^{n]}
  +\wt\psi^{[m}\wt\psi^{n]}).
\end{equation}
By substituting the solution (\ref{freesol}) to (\ref{Lint3}),
we obtain the following expression with the help of the Bianchi identity.
\begin{eqnarray}
H_{\rm int}
&=&-\frac{\alpha'}{4}\int_0^{2\pi}\frac{d\sigma}{2\pi}
R_{mnpq}K^{mn}\wt K^{pq}+H'\nonumber\\
&=&-\frac{\alpha'}{4}\sum_k
R_{mnpq}K_k^{mn}\wt K_k^{pq}+H',
\end{eqnarray}
where
$K_k^{mn}$ and $\wt K_k^{mn}$ are Fourier modes of $K^{mn}$ and $\wt K^{mn}$.
$H'$ is defined by
\begin{eqnarray}
H'
&=&\int_0^{2\pi}\frac{d\sigma}{2\pi\alpha'}\Big[
-\frac{i}{2}R_{pqmn}(\wt\phi^p\wt\psi^m\wt\psi^n)(\partial_+\phi^q)
-\frac{i}{2}R_{pqmn}(\phi^p\psi^m\psi^n)(\partial_-\wt\phi^q)
\nonumber\\
&&
+\frac{2}{3}
R_{mpnq}(\wt\phi^p\wt\phi^q\partial_-\wt\phi^n)(\partial_+\phi^m)
+\frac{2}{3}
R_{mpnq}(\phi^p\phi^q\partial_+\phi^m)(\partial_-\wt\phi^n)
\nonumber\\
&&
+\frac{1}{6}
R_{mpqn}\partial_+(\phi^m\phi^n)\partial_-(\wt\phi^p\wt\phi^q)
\Big].
\end{eqnarray}
All the terms in $H'$ are in fact irrelevant to our computation of
the mass correction by the following reason.
Let us look at the first term in $H'$.
It is a product of the left moving part
$\partial_+\phi^q$ and the right moving part $\wt\phi^p\wt\psi^m\wt\psi^n$.
We are interested in the leading term in the
perturbative expansion of the mass correction.
At this level of perturbation,
we need only the energy-preserving matrix elements of the interaction
operator $H_{\rm int}$.
However, because of the derivative in the left moving factor $\partial_+\phi^q$,
this term does not have energy-preserving matrix elements.
This is the case for any term in $H'$.
Thus $H'$ is irrelevant to our computation.

Furthermore, because both the operators $K_k^{mn}$ and $\wt K_k^{mn}$ change the
worldsheet energy $L_0+\wt L_0$ by $k$, only the zero mode with $k=0$
has energy preserving parts
contributing to the mass correction.

Finally we obtain a quite simple formula for the interaction term
relevant to the mass correction.
\begin{equation}
H_{\rm int}=-\frac{\alpha'}{4} R_{mnpq}K_0^{mn}\wt K_0^{pq}.
\label{HRKK}
\end{equation}
Note that the Fourier index $k$ of $K_k^{mn}$ and $\wt K_k^{mn}$
is not always an integer
because the currents $J^{mn}$ satisfy the twisted boundary condition
$J^{mn}(\sigma+2\pi)=UJ^{mn}(\sigma)U^{-1}$.
Only generators commuting with $U$ have zero modes
and contribute to the mass correction.

\section{Deformation of supersymmetric orbifolds}\label{susy.sec}
In this and the next sections, we use the formula (\ref{HRKK})
to compute the mass corrections to the twisted modes
in several orbifolds.
Although the formula is quite general and we can in principle
compute the mass correction for any states in twisted sectors,
we focus on low-lying states in the NS-NS sector
because we are interested in the geometric instability of
the orbifold.

In this section, we discuss deformation of supersymmetric orbifolds.

\subsection{${\bf C}^2/{\bf Z}_{N(1)}$}\label{czn1.sec}
First, we consider supersymmetric ${\bf C}^2/{\bf Z}_{N(1)}$ orbifold.
Before computing the mass corrections,
let us summarize the low-lying spectrum of the flat orbifold
without deformation.

In four dimension, the rotational symmetry is
$\Spin(4)=\SU(2)_L\times\SU(2)_R$, and it
is convenient to use undotted ($\SU(2)_L$) and dotted ($\SU(2)_R$) spinor indices instead of vector ones.
These indices are transformed to each other with matrices
$(\sigma^m)^a{}_{\dot b}$ and their hermitian conjugate $(\ol\sigma^m)^{\dot a}{}_b$.
The $\Spin(4)$ currents $J^{mn}$ are decomposed into $\SU(2)_L$ and
$\SU(2)_R$ parts with two spinor indices by
\begin{equation}
\frac{1}{2}J_{mn}(\sigma^m)^a{}_{\dot b}(\ol\sigma^n)^{\dot a}{}_b=J^a{}_b\delta^{\dot a}_{\dot b}+J^{\dot a}{}_{\dot b}\delta^a_b.
\end{equation}
With this definition,
the current with spinor indices $J^{12}$ is
a hermitian operator
and eigenvalues of its zero-mode $J_0^{12}$
for $\SU(2)_L$ doublet are $\pm1/2$.
The $\SU(2)_R$ current $J^{\dot1\dot2}$ is also normalized in the same way.
The target space coordinates $\Phi^m$ and their superpartners $\Psi^m$
are also rewritten as bi-spinors $\Phi^{a\dot a}$ and $\Psi^{a\dot a}$.
We consider ${\bf C}^2/{\bf Z}_N$ orbifold
with the ${\bf Z}_N$ generator $\Gamma$ defined by
\begin{equation}
\Gamma=\exp\left(\frac{4\pi i}{N}(J_0^{\dot1\dot2})\right).
\label{susygamma}
\end{equation}
In this orbifold, half of supersymmetries belonging to ${\bf 2}_R$ are broken
and the other half in ${\bf 2}_L$ remains unbroken.
(${\bf2}_L$ and ${\bf2}_R$ denote $({\bf2},{\bf1})$ and $({\bf1},{\bf2})$
representations of $\SU(2)_L\times\SU(2)_R$, respectively.)
There are $N-1$ twisted sectors labeled
by a non-zero ${\bf Z}_N$ integer $w$.
We also use a parameter $\kappa=w/N$ instead of $w$.
We consider the NS-NS sector of a twisted mode with boundary condition
\begin{equation}
\Phi^{a\dot1}(\sigma+2\pi)=e^{2\pi i\kappa}\Phi^{a\dot1}(\sigma),\quad
\Psi^{a\dot1}(\sigma+2\pi)=-e^{2\pi i\kappa}\Psi^{a\dot1}(\sigma).
\end{equation}
$\Phi^{a\dot2}$ and $\Psi^{a\dot2}$ are expressed with $\Phi^{a\dot1}$
and $\Psi^{a\dot1}$ by the hermiticity conditions
$\epsilon_{ac}\epsilon_{\dot b\dot d}\Phi^{c\dot d}=(\Phi^{a\dot b})^\dagger$
and
$\epsilon_{ac}\epsilon_{\dot b\dot d}\Psi^{c\dot d}=(\Psi^{a\dot b})^\dagger$.
This theory can be quantized in the ordinary way in the light-cone formalism.
Let us assume $0<\kappa\leq1/2$.
The free part of the hamiltonian $H_0=L_0+\wt L_0$ is given by
\begin{equation}
L_0=N-\frac{1}{2}+\kappa,\quad
\wt L_0=\wt N-\frac{1}{2}+\kappa.
\end{equation}
We are interested only in massless and tachyonic states.
To make such states, only creation oscillators we can act on the vacuum state
$|0\rangle_L\times|0\rangle_R$ are
the following four.
\begin{equation}
\psi_{-1/2+\kappa}^{a\dot 2},\quad
\wt\psi_{-1/2+\kappa}^{a\dot 1},\quad
\phi_{-\kappa}^{a\dot 1},\quad
\wt\phi_{-\kappa}^{a\dot 2}.
\label{kisnt12}
\end{equation}
We have additional fermionic zero modes
for $\kappa=1/2$. In the supersymmetric case, however,
we can obtain massless spectrum in the $\kappa=1/2$ case
as a limit of $\kappa<1/2$ case
and we do not have to treat it separately.
This is not the case for non-supersymmetric orbifold we will discuss later.

In the case of supersymmetric orbifold,
both $|0\rangle_L$ and $|0\rangle_R$ are GSO odd,
and we need at least one fermionic oscillators
in both the left and the right moving parts.
We have four degenerate massless states.
\begin{equation}
|ab\rangle\equiv \psi^{a\dot2}_{-1/2+\kappa}|0\rangle_L\otimes
\wt \psi^{b\dot1}_{-1/2+\kappa}|0\rangle_R
\end{equation}
These states carry $K_0^{\dot1\dot2}=\wt K_0^{\dot1\dot2}=0$ and are invariant under $\Gamma$.
We recombine $|ab\rangle$ into the $\SU(2)_L$ singlet $|{\bf 1}\rangle$
and the triplet $|{\bf 3}(m)\rangle$ where $m=0,\pm1$ represents the
$J^{12}$ eigenvalue.
The triplet states for each $w$ corresponds to blow-up modes
associated with one of ${\bf S}^2$ cycles of $A_{N-1}$ singularity,
while the singlet mode for each $w$ is related to the $B$-field
integrated over each two cycle.

Let us now compute the mass correction to each state we have obtained above.
In four dimension, the Weyl tensor has $10$ independent components
belonging to ${\bf 5}_L+{\bf 5}_R$.
This can be easily checked as follows.
Because we can regard each pair of indices $mn$ and $pq$ of the curvature tensor
$R_{mnpq}$ as an index for the adjoint representation of $\Spin(4)=\SU(2)_L\times \SU(2)_R$, $R_{mnpq}$ belongs to
the symmetric product of two adjoint representations
$[({\bf 3}_L+{\bf 3}_R)\times({\bf 3}_L+{\bf 3}_R)]_{\rm sym}
={\bf5}_L+{\bf5}_R+{\bf3}_L\times{\bf3}_R+{\bf1}+{\bf1}$.
One of the singlets
and ${\bf 3}_L\times {\bf 3}_R$
correspond to the ten components of the Ricci tensor $R_{mn}$,
which is assumed to vanish.
The other singlet represents $\epsilon^{mnpq}R_{mnpq}$,
which also vanishes owing to the Bianchi identity.
Thus, we have only ten independent components of the curvature tensor
belonging to ${\bf 5}_L+{\bf 5}_R$.
We can extract ${\bf 5}_L$ components $R_{abcd}$ and ${\bf 5}_R$ components
$R_{\dot a\dot b\dot c\dot d}$ as
\begin{equation}
\frac{1}{4}
R_{mnpq}
(\sigma^m)^a{}_{\dot a}
(\ol\sigma^n)^{\dot b}{}_b
(\sigma^p)^c{}_{\dot c}
(\ol\sigma^q)^{\dot d}{}_d
=R^a{}_b{}^c{}_d\delta_{\dot a}^{\dot b}\delta_{\dot c}^{\dot d}
+R_{\dot a}{}^{\dot b}{}_{\dot c}{}^{\dot d}\delta^a_b\delta^c_d.
\end{equation}
Both $R_{abcd}$ and $R_{\dot a\dot b\dot c\dot d}$ are
completely symmetric in all of their indices.
In terms of these tensors with spinor indices,
the interaction term (\ref{HRKK}) is rewritten as
\begin{equation}
H_{\rm int}=
-\alpha' R_{abcd}K_0^{ab}\wt K_0^{cd}
-\alpha' R_{\dot a\dot b\dot c\dot d}K_0^{\dot a\dot b}\wt K_0^{\dot c\dot d}.
\label{HRR}
\end{equation}

If we turn on only the $R_{\dot a\dot b\dot c\dot d}$ with keeping $R_{abcd}=0$,
the $\SU(2)_L$ part of the holonomy remains trivial.
This implies that this deformation does not break supersymmetry.
Therefore, we expect the states $|ab\rangle$ remain massless.
Indeed, because the second term on the right hand side in (\ref{HRR})
consists of $\SU(2)_R$ generators,
which act trivially on the massless states $|ab\rangle$,
all the $H_{\rm int}$ eigenvalues for $|ab\rangle$ are zero.

On the other hand, if we turn on the ${\bf 5}_L$ part of the curvature $R_{abcd}$,
the $\SU(2)_L$ holonomy becomes nontrivial and supersymmetry is completely broken.
Although we can compute correction in general case,
let us consider deformation keeping $\U(1)_L$ isometry for simplicity.
If we impose this isometry on the deformation,
only the component $R_{1122}$ of the curvature tensor is allowed
to be non-zero.
If it is negative, we can extend this deformation
to the whole ${\bf C}^2$ such that the space becomes Taub-NUT manifold.
The relation between the asymptotic radius $R_{\rm TN}$
of the Taub-NUT manifold and the curvature at the center is
\begin{equation}
R_{1122}=-\frac{4}{R_{\rm TN}^2}.
\end{equation}
The interaction induced by this deformation is
\begin{equation}
H_{\rm int}=
\frac{4\alpha'}{R_{\rm TN}^2}(4K_0^{12}\wt K_0^{12}+K_0^{11}\wt K_0^{22}+K_0^{22}\wt K_0^{11})
\label{HRRtn}
\end{equation}
We can easily see that $H_{\rm int}$ is diagonal
on states $|{\bf 1}\rangle$ and $|{\bf3}(m)\rangle$.
Because $H_{\rm int}$ is written in terms of the $\SU(2)_L$ generators,
it trivially vanishes on the singlet state $|{\bf1}\rangle$.
\begin{equation}
H_{\rm int}|{\bf1}\rangle=0,
\label{eq20}
\end{equation}
As we mentioned before,
$|{\bf 1}\rangle$ corresponds to the
$B$-field integrated over a two-cycle.
We can interpret (\ref{eq20}) as a result of
the $B$-field gauge invariance.
The eigenvalue of $H_{\rm int}\sim(\alpha'/2)\delta m^2$ on each state in the triplet is easily obtained
as
\begin{equation}
H_{\rm int}|{\bf3}(0)\rangle=-\frac{8\alpha'}{R_{\rm TN}^2}|{\bf3}(0)\rangle,\quad
H_{\rm int}|{\bf3}(\pm1)\rangle=\frac{4\alpha'}{R_{\rm TN}^2}|{\bf3}(\pm1)\rangle
\end{equation}
Therefore, only one of the triplet modes becomes tachyonic.


\subsection{${\bf C}^3/{\bf Z}_3$ and ${\bf C}^4/{\bf Z}_4$}
In the same way as the four dimensional case, 
we can compute the mass correction for other dimensions.
Here we explicitly calculate the
mass corrections to massless twisted modes of
supersymmetric orbifolds
${\bf C}^3/{\bf Z}_3$ and ${\bf C}^4/{\bf Z}_4$.

We choose the complex coordinates ($z^1,z^2,z^3$) of ${\bf C}^3$ 
to be $z^1=x^1+ix^2$, $z^2=x^3+ix^4$, $z^3=x^5+ix^6$. 
The orbifold is constructed by dividing this ${\bf C}^3$ by 
${\bf Z}_3$ action.
The generator $\Gamma$ of
${\bf Z}_3$ is expressed as 
\begin{equation}
\Gamma=
(-1)^{F_s}{\rm exp}\left(
\frac{2\pi}{3}J_0^{12}+\frac{2\pi}{3}J_0^{34}+\frac{2\pi}{3}J_0^{56}
\right),
\end{equation}
where $F_s$ is the spacetime fermion number.
There are two twisted sectors corresponding to
nontrivial elements $\Gamma$ and $\Gamma^2$ of ${\bf Z}_3$,
and we have one massless state in each sector.
We denote them by $|{\bf1}\rangle_\Gamma$ and $|{\bf1}\rangle_{\Gamma^2}$.

In six dimension the curvature tensor belongs to 
${\bf 84}_{(202)}+{\bf 20}_{(020)}+{\bf 1}_{(000)}$ 
representation of $\SO(6)\simeq\SU(4)$. 
(Here we attached the Dynkin indices to distinguish same dimensional  representation.) 
The Ricci-flat condition removes ${\bf 20}_{(020)}+{\bf 1}_{(000)}$
and there remains ${\bf 84}_{(202)}$.
This ${\bf 84}_{(202)}$ representation of $\SU(4)$ decomposes into representations of $\SU(3)$, and among them we can use 
${\bf 27}_{(22)}+{\bf 8}_{(11)}+{\bf 1}_{(00)}$ 
as the ${\bf Z}_3$ invariant deformations. 
Therefore the curvature tensor
can be written explicitly as
\begin{align}
R_{i\bar\jmath k\bar l}
&=R_{i\bar\jmath k\bar l}^{(27)}
+R_{i\bar\jmath k\bar l}^{(8)}
+R_{i\bar\jmath k\bar l}^{(1)}\nonumber\\
&=T_{ik\bar\jmath\bar l}^{(27)}
+[4(T_{i\bar\jmath}^{(8)}\delta_{k\bar l}
+T_{k\bar l}^{(8)}\delta_{i\bar\jmath})
-(T_{i\bar l}^{(8)}\delta_{k\bar j}
+T_{k\bar\jmath}^{(8)}\delta_{i\bar l})]\nonumber\\
&+T^{(1)}[5\delta_{i\bar\jmath}\delta_{k\bar l}
+\delta_{i\bar l}\delta_{k\bar\jmath}],\label{R_C3}
\end{align}
where $i,j,\ldots$ ($\bar\imath,\bar\jmath,\ldots$)
denote holomorphic (anti-holomorphic) 
indices of ${\bf 3}_{(10)}$ ($\bar{\bf 3}_{(01)}$) representation.
The tensor $T_{ij\bar k\bar l}^{(27)}$
belongs to
${\bf 27}_{(22)}$ and satisfies
\begin{equation}
T_{ij\bar k\bar l}^{(27)}
=T_{ji\bar k\bar l}^{(27)}
=T_{ij\bar l\bar k}^{(27)},\quad
T_{ij\bar k\bar l}^{(27)}\delta^{i\bar k}=0.
\label{T27}
\end{equation}
Similarly $T_{i\bar\jmath}^{(8)}$ is ${\bf 8}_{(11)}$ representation
tensor satisfying $T_{i\bar\jmath}^{(8)}\delta^{i\bar\jmath}=0$. 
$T^{(1)}$ represents the singlet degree of freedom.
Using Bianchi identity, we can calculate $R_{ik\bar\jmath\bar l}$ 
component from \eqref{R_C3}, {\it i.e.}
$R_{ik\bar\jmath\bar l}=-R_{i\bar\jmath\bar lk}
-R_{i\bar lk\bar\jmath}
$. 
We can determine whether the deformation preserves K\"ahler 
property or not depending on whether $R_{ik\bar\jmath\bar l}$ 
equals zero or not.
Actually we can see that ${\bf 27}_{(22)}$ deformation preserves 
K\"ahler property.
On the other hand an orbifold deformed by 
${\bf 8}_{(11)}$ or ${\bf 1}_{(00)}$ components is not K\"ahler
and all the supersymmetries are broken in it.

As we have already obtained the explicit formula \eqref{HRKK} 
for general deformations, 
we can easily calculate the mass corrections to
$|{\bf1}\rangle _\Gamma$ and $|{\bf1}\rangle _{\Gamma^2}$.
Because these massless states are $\SU(3)$ singlets,
non-zero mass corrections come from the trace part
of the generators $K^{l\bar m}$ and ${\wt K}^{l\bar m}$:
\begin{align}
\langle K^{l\bar m}\rangle _{\Gamma}
=\langle \wt K^{l\bar m}\rangle _{\Gamma^2}= \delta^{l\bar m},\quad
\langle \wt K^{l\bar m}\rangle _{\Gamma}
=\langle K^{l\bar m}\rangle _{\Gamma^2}&= -\delta^{l\bar m}.
\end{align}
As a result we obtain the following mass correction to the massless states: 
\begin{equation}
\frac{\alpha'}{2}m^2
=H_{\rm int} = 48\alpha' T^{(1)}.
\label{6dm}
\end{equation}

For ${\bf C}^4/{\bf Z}_4$ orbifold, 
we choose the complex coordinates in the similar way as 
${\bf C^3}$ case ($z^1=x^1+ix^2,\cdots$) and the action of the generator 
$\Gamma$ to be 
\begin{equation}
\Gamma=
(-1)^{F_s}
{\rm exp}\left(
\frac{2\pi}{4}J_0^{12}+\frac{2\pi}{4}J_0^{34}+
\frac{2\pi}{4}J_0^{56}+\frac{2\pi}{4}J_0^{78}
\right),
\end{equation}
so that this orbifold is supersymmetric. 
There are two massless states in $\Gamma$ and $\Gamma^3$ 
twisted sectors.
(All the $\Gamma^2$ twisted modes are massive. ) 
The eight dimensional curvature tensor belongs to 
${\bf 300}_{(000;2)}+{\bf 35}_{(200;0)}+{\bf 1}_{(000;0)}$ 
representation of $\SO(8)$. (Here we used the convention 
in which the first three Dynkin indices are related by triality.) 
Among them, Ricci-flat curvature tensor belongs to 
${\bf 300}_{(000;2)}$. 
Decomposing ${\bf 300}_{(000;2)}$ into $\SU(4)$ representation and 
picking up ${\bf Z}_4$ invariant ones, we obtain
${\bf 84}_{(202)}+{\bf 20}_{(020)}+{\bf 15}_{(101)}+{\bf 1}_{(000)}$. 
The curvature tensor can be written as 
\begin{align}
R_{i\bar\jmath k\bar l}
&=R_{i\bar\jmath k\bar l}^{(84)}+
R_{i\bar\jmath k\bar l}^{(20)}+
R_{i\bar\jmath k\bar l}^{(15)}+
R_{i\bar\jmath k\bar l}^{(1)}\nonumber\\
&=T_{ik\bar\jmath\bar l}^{(84)}
+T_{ik\bar\jmath\bar l}^{(20)}
+[T_{i\bar\jmath}^{(15)}\delta_{k\bar l}
+T_{k\bar l}^{(15)}\delta_{i\bar\jmath}]
+T^{(1)}[7\delta_{i\bar\jmath}\delta_{k\bar l}
+2\delta_{i\bar l}\delta_{k\bar\jmath}].
\end{align}
$T_{ij\bar k\bar l}^{(84)}$ satisfies the same condition 
as $T_{ij\bar k\bar l}^{(27)}$ in \eqref{T27}. 
$T_{ij\bar k\bar l}^{(20)}$ satisfies the condition 
$
T_{ij\bar k\bar l}^{(20)}
=-T_{ji\bar k\bar l}^{(20)}
=-T_{ij\bar l\bar k}^{(20)}$ and
$T_{ij\bar k\bar l}^{(20)}\delta^{i\bar k}=0$.
$T_{i\bar\jmath}^{(15)}$ tensor satisfies the same condition 
as $T_{i\bar\jmath}^{(8)}$ in the six dimensional case;
$T_{i\bar\jmath}^{(15)}\delta^{i\bar\jmath}=0$. 
In the same way as the six dimensional case, we can determine 
whether each deformation preserves K\"ahler property. 
Actually only ${\bf 84}_{(202)}$ deformation preserves 
K\"ahler property and other three deformations lead to 
non-K\"ahler manifolds. 

Mass correction of massless states caused by these deformation
are also easily calculated in the same way 
as the six dimensional case, and we obtain
\begin{equation}
\frac{\alpha'}{2}m^2=
H_{\rm int} = 120\alpha' T^{(1)}.
\label{8dm}
\end{equation}

In both six and eight dimensional cases,
the mass corrections (\ref{6dm}) and (\ref{8dm})
depend only on the singlet component of the curvature,
which break all the supersymmetries.
We can choose, at least locally, the signature of the deformation parameters
so that these states become tachyonic, although it is not clear if we can extend these deformations to the whole space.

\section{Non-supersymmetric ${\bf C}^2/{\bf Z}_{N(N+1)}$}\label{nonsusy.sec}
As is mentioned in the introduction,
one purpose of this paper is to confirm the
result in \cite{unstablens5},
in which the closed string tachyon condensation
on a type 0 NS5-brane
in ${\bf S}^1$ compactified background is
identified with decay of spacetime with creating a Witten's bubble.
In section 4.1 we analyze the tachyonic masses including
numerical factors in the type 0 side.
However, as we will comment later,
calculations in section 4.1 are actually quite
unreliable
because of the large curvature of type 0 settings and
the smearing of NS5-branes.
The reason why we present this analysis is to give an intuitive explanation
for the existence of localized tachyons on type 0 NS5-branes.
In 4.2 we give a more reliable computation using the formula (\ref{HRKK})
for the deformed orbifold and then we will see agreements
except some corrections
due to the localization of NS5-branes.

\subsection{Tachyonic modes on type 0 NS5-branes}\label{KG.sec}
Let us consider $N_{\rm NS5}$ parallel type 0 NS5-branes.
Type 0 theory has the closed string tachyon field in the bulk.
To localize it on NS5-branes,
we compactify one direction transverse to the NS5-branes,
and impose anti-periodic boundary condition
on the tachyon field.
If the compactification radius is sufficiently small,
the bulk tachyon becomes massive
due to the large Kaluza-Klein momentum.
However, there are remnant of tachyonic modes near the NS5-branes
as we see below \cite{imamura}.
We are interested in the T-duality
between these NS5-branes and an orbifold.
More precisely, NS5-branes dual to the orbifold we will discuss later
are distributed along
an ${\bf S}^1$ at even intervals.
Furthermore, at the supergravity level, the dual configuration
is the smeared NS5-brane solution
\begin{equation}
ds^2=\eta_{ij}dx^idx^j+H(x^\mu)\delta_{\mu\nu}dx^\mu dx^\nu,\quad
e^{2\phi}=\mbox{const.}\times H(x^\mu),
\label{NS5solution}
\end{equation}
where $x^i$ ($i=0,\ldots,5$) and $x^\mu$ ($\mu=6,7,8,9$) are
parallel and transverse coordinates respectively.
$x^9$ is periodic coordinate with period $2\pi$.
The harmonic function $H$ is given by
\begin{equation}
H=\frac{\alpha'N_{\rm NS5}}{2r}+R^2,\quad
r^2=(x^6)^2+(x^7)^2+(x^8)^2.
\end{equation}
The parameter $R$ determine the ${\bf S}^1$ radius
at the asymptotic region $r\rightarrow\infty$.
When we regard this as the T-dual of the orbifold,
we should set $R=0$.
Later we discuss the deformation of orbifold to
a Kaluza-Klein monopole geometry with asymptotic radius $\wt R$.
The Kaluza-Klein monopole is described as a ${\bf Z}_{2N_{\rm NS5}}$
orbifold of the Taub-NUT
manifold with asymptotic radius $R_{\rm TN}$.
The relation among these radii is
\begin{equation}
\frac{R_{\rm TN}}{2N_{\rm NS5}}=\wt R=\frac{\alpha'}{2R}.
\end{equation}
We can obtain localized tachyons on NS5-branes
as normalizable solutions of the equation of motion
\begin{equation}
\partial_ig^{ij}\partial_j T
=\left(-\frac{e^{2\phi}}{\sqrt{-g}}\partial_\mu g^{\mu\nu}\frac{\sqrt{-g}}{e^{2\phi}}\partial_\nu
+M_T^2\right)T.
\label{KGeq}
\end{equation}
(When we obtain this equation of motion, it is important that
$(H_{\mu\nu\rho})^2T^2$ term is absent in the action of type 0 theory \cite{garousi}.)
Therefore, the problem determining mass spectrum of localized tachyons
on the NS5-branes reduces to
one determining
eigenvalue of the four-dimensional Klein-Gordon like operator on the right hand side in (\ref{KGeq}).
Each eigen function factorizes into an exponential factor $e^{ikx^9}$
associated with ${\bf S}^1$ direction and function $\psi(x^6,x^7,x^8)$
depending on three non-compact transverse coordinates.
Because the tachyon field is anti-periodic
along ${\bf S}^1$, $k$ is a half odd integer.
The differential equation satisfied by $\psi$ has the same structure with
the Schr\"odinger equation for a particle in a Coulomb potential,
and is easily solved analytically.
The wave functions and corresponding eigenvalues
are labeled by four quantum numbers.
One is the Kaluza-Klein momentum $k$ along ${\bf S}^1$.
The other three are the same with those for a hydrogen atom.
Namely, the principal quantum number $n$,
the azimuthal one $l$ and the magnetic one $m$.
By solving (\ref{KGeq}),
we obtain the following mass spectrum
\begin{equation}
\frac{\alpha'}{2}M_6^2
=
-1+\frac{2n|k|}{N_{\rm NS5}}
-\frac{4n^2\alpha'}{R_{\rm TN}^2}+{\cal O}(\alpha'/R_{\rm TN}^4),
\label{m6bywtR}
\end{equation}
where the normalization is so chosen that this can be identified with
worldsheet hamiltonian $L_0+\wt L_0$.
Just as the spectrum of a hydrogen atom, the mass eigenvalues do not depend on
$l$ and $m$, and for each $k$ and $n$, there are $n^2$ degenerate states.

For $N_{\rm NS5}=1$, the corresponding ${\bf C}^2/{\bf Z}_2$ orbifold is supersymmetric and
there is no tachyonic state.
Actually, the first two terms on the right hand side in (\ref{m6bywtR})
cancels each other for $(k,n)=(\pm1/2,1)$ (1s state).
If it is deformed and $R_{\rm TN}$ becomes finite,
the leading mass correction is given by the third term in (\ref{m6bywtR}).
The tachyonic mass does not include string scale.
(The common $\alpha'$ factors on the left and right hand sides can be removed.)
This fact suggests us that we may be able to analyze the dual configuration
by means of classical gravity.
Indeed, it was shown in \cite{unstablens5} that
we reproduce the same tachyonic mass up to numerical factor
by analyzing the unstable mode associated with the creation of Witten's bubble.
In \cite{unstablens5}, only the existence of the unstable modes
is shown, but
the number of unstable modes and
numerical values of the tachyonic masses
are left undetermined.
In the next section, we give the answer to this question
by computing the mass correction to the
twisted modes.

Before ending this subsection,
we would like to emphasize the following point.
Here we treated the tachyon field quantum mechanically, and neglected
other stringy excitations.
In fact this cannot be justified because the curvature of the
background spacetime is of the order of the string scale.
Furthermore, we use a smeared solution rather than a localized one.
Even if the parameter $R$ is however small,
using a smeared solution is not a good approximation because the
proper compactification radius near the NS5-branes is not small
due to the divergent metric.

In the next subsection, we reconsider the tachyonic spectrum in the
dualized type II theory.
There the calculations are more reliable and
we obtain the correct tachyonic mass spectrum

\subsection{Twisted mode analysis}\label{twisted.sec}
The dual configuration to the type 0 NS5-branes discussed above
is Taub-NUT deformation of non-supersymmetric ${\bf C}^2/{\bf Z}_{N(N+1)}$ orbifold
with $N=2N_{\rm NS5}$ \cite{BG,ima1}.
Note that $N$ is always even integer.
It can be realized by replacing the
generator $\Gamma$ in (\ref{susygamma}) for the supersymmetric orbifold
by
\begin{equation}
\Gamma=(-)^{F_s}\exp\left(\frac{4\pi i}{N}(J^{\dot1\dot2}_0)\right).
\label{nonsusygamma}
\end{equation}
The sectors with even $w$ are identical to the supersymmetric case
which we have already discussed in the last section.
We only need to consider odd $w$ sectors separately.
The extra factor $(-1)^{F_s}$ in (\ref{nonsusygamma}) reverses the GSO parity
in the odd $w$ sectors,
and both $|0\rangle_L$ and $|0\rangle_R$ become GSO even.
We have to act even number of fermionic oscillators on
the vacuum state in each of the left and the right moving parts.

Let us consider $0<\kappa<1/2$ case first and leave the $\kappa=1/2$ case
for later.
If $0<\kappa<1/2$, the excitation by two fermionic oscillators makes the state massive.
Therefore, we can only use bosonic ones in (\ref{kisnt12}).
\begin{equation}
|n\rangle\equiv \phi_{-\kappa}^{a_1\dot1}
\cdots \phi_{-\kappa}^{a_{n-1}\dot1}
|0\rangle_L\otimes
\wt \phi_{-\kappa}^{b_1\dot2}
\cdots\wt \phi_{-\kappa}^{b_{n-1}\dot2}
|0\rangle_R
\label{nstate}
\end{equation}
Each of left and right moving part
has $n-1$ symmetric $\SU(2)_L$ indices and thus belongs to spin $(n-1)/2$
representation of $\SU(2)_L$.
Combining the left and right moving parts,
we obtain $n^2$ degenerate states.
These degenerate states together are denoted by $|n\rangle$ and we
will not distinguish each of them.
The $\U(1)_R$ charges of these states are given by
\begin{equation}
K_0^{\dot 1\dot 2}|n\rangle=\frac{n}{2}|n\rangle,\quad
\wt K_0^{\dot 1\dot 2}|n\rangle=-\frac{n}{2}|n\rangle.
\label{u1charge}
\end{equation}
Therefore, $J_0^{\dot1\dot2}|n\rangle=0$ and
these states are invariant under $\Gamma$.
The energy of these states are
\begin{equation}
(L_0+\wt L_0)|n\rangle=(-1+2n\kappa)|n\rangle.
\end{equation}
The states $|n\rangle$ 
are tachyonic if $n$ is smaller than $1/(2\kappa)$.
This reproduce the spectrum in (\ref{m6bywtR}) with $R_{\rm TN}\rightarrow\infty$.
(Note that $\kappa=w/N$ is identified with $k/N_{\rm NS5}$.)

Let us consider deformation of this orbifold.
We restrict our attention to the Taub-NUT deformation
which is dual to the type 0 NS5-branes.
This deformation keeps the $\SU(2)_L\times\U(1)_R$ isometry,
and
only non-vanishing component of the curvature tensor
is $R_{\dot1\dot1\dot2\dot2}=-4/R_{\rm TN}^2$,
where $R_{\rm TN}$ is the asymptotic radius of the Taub-NUT manifold.
When $\kappa\neq1/2$,
currents
$J^{\dot 1\dot 1}$ and $J^{\dot 2\dot 2}$
do not commute with
$\Gamma^w\in\U(1)_R$ and their zero modes are absent.
The Hamiltonian is simplified to
\begin{equation}
H_{\rm int}=-4\alpha'R_{\dot 1\dot 1\dot 2\dot 2}K_0^{\dot 1\dot 2}\wt K_0^{\dot 1\dot 2},
\label{genericcor}
\end{equation}
and the mass correction
to $|n\rangle$ is
easily obtained as
\begin{equation}
H_{\rm int}|n\rangle
=\alpha'n^2 R_{\dot1\dot1\dot2\dot2}|n\rangle
=-\frac{4n^2\alpha'}{R_{\rm TN}^2}|n\rangle.
\end{equation}
This precisely reproduce the third term in (\ref{m6bywtR}),
obtained by solving the Klein-Gordon equation
on the smeared NS5-brane solution.

In the $\kappa=1/2$ case, which occurs if $N=2\mod4$, there are additional massless states.
Let us look at
a state
$\psi^{1\dot2}_{-1/2+\kappa}\psi^{2\dot2}_{-1/2+\kappa}|0\rangle_L$
in the left moving part.
The energy of this state $L_0=1/2-\kappa$
vanishes when $\kappa=1/2$.
This massless state makes $\SU(2)_R$ doublet together with another
massless state $|0\rangle_L$.
Therefore, combining with right moving part, which is also $\SU(2)_R$ doublet
in the same manner,
we have four states belonging to ${\bf2}_R\times{\bf2}_R$.
We decompose them to the $\SU(2)_R$ singlet $|\dot{\bf1}\rangle$ and the $\SU(2)_R$ triplet $|\dot{\bf3}(m)\rangle$ just as in the supersymmetric case.
In the $N=2$ case, the orbifold ${\bf C}^2/{\bf Z}_{2(2+1)}$ is accidentally
supersymmetric, and is the parity transformation of ${\bf C}^2/{\bf Z}_{2(1)}$, which we studied in the previous section.
We can identify $|\dot{\bf 3}(m)\rangle$ and $|\dot{\bf1}\rangle$
with the blow-up modes
of $A_1$ singularity and the $B$-field integrated over the $2$-cycle,
respectively.

For the $\kappa=1/2$ sector, $\Spin(4)$ symmetry is recovered and,
in addition to $K^{\dot1\dot2}$ and $\wt K^{\dot1\dot2}$,
four charges
$K_0^{\dot1\dot1}$, $K_0^{\dot2\dot2}$,
$\wt K_0^{\dot1\dot1}$, and $\wt K_0^{\dot2\dot2}$ contribute to the
mass correction.
\begin{equation}
H_{\rm int}=-\alpha'R_{\dot 1\dot 1\dot 2\dot 2}
(4K_0^{\dot 1\dot 2}\wt K_0^{\dot 1\dot 2}
 +K_0^{\dot 1\dot 1}\wt K_0^{\dot 2\dot 2}
 +K_0^{\dot 2\dot 2}\wt K_0^{\dot 1\dot 1})
\label{k12case}
\end{equation}
The matrix elements of $H_{\rm int}$ can be computed in the same way with ${\bf 5}_L$ deformation
of the supersymmetric orbifold.
\begin{equation}
H_{\rm int}|\dot{\bf1}\rangle=0,\quad
H_{\rm int}|\dot{\bf3}(0)\rangle=-\frac{8\alpha'}{R_{\rm TN}^2}|\dot{\bf3}(0)\rangle,\quad
H_{\rm int}|\dot{\bf3}(\pm1)\rangle=\frac{4\alpha'}{R_{\rm TN}^2}|\dot{\bf3}(\pm1)\rangle
\label{eq45}
\end{equation}
We can understand the masslessness of the singlet state
as a result of the $B$-field gauge symmetry.
There is only one tachyonic state, which carries
no $\U(1)_L\times\U(1)_R$ charges,
and it is nothing but the unstable mode associated with the creation
of a Witten's bubble discussed in
\cite{unstablens5}.
We have now confirmed that this is the unique unstable mode
and obtain the mass including the numerical factor,
which was not obtained in \cite{unstablens5}.

Let us compare the mass correction (\ref{eq45}) obtained here to (\ref{m6bywtR}).
In the previous subsection, we only considered Kaluza-Klein modes
without windings.
Therefore, we should compare them to twisted modes
without Kaluza-Klein momentum
$J_0^{\dot1\dot2}=K_0^{\dot1\dot2}+\wt K_0^{\dot1\dot2}$.
Among four states, we have two such states $|\dot{\bf1}\rangle$ and
$|\dot{\bf3}(0)\rangle$.
One is massless and the other is tachyonic.
On the other hand, from the equation (\ref{m6bywtR}), we expect the emergence of two
degenerate tachyonic modes with quantum numbers $(k,n)=(\pm1/2,1)$.
This discrepancy is explained as follows.
We used the smeared NS5-brane solution in \S\ref{KG.sec} to obtain (\ref{m6bywtR}).
It gives only momentum $P$ preserving amplitude correctly.
The NS5-brane is actually localized at a point on the circle,
and it causes $P$-violating scattering.
The momentum operator $P$ in the type 0 side
is related to $\Spin(4)$ charges
by $P=K_0^{\dot 1\dot 2}-\wt K_0^{\dot 1\dot2}$,
which does not commute with $H_{\rm int}$ in (\ref{k12case}).
If we ignore the $P$-violating terms in (\ref{k12case}),
the interaction Hamiltonian reduces to (\ref{genericcor}), which we used
for $\kappa\neq1/2$,
and we obtain the degenerate tachyonic mass obtained
in \S\ref{KG.sec}.
Therefore, we conclude that the discrepancy between
Klein-Gordon equation analysis and the twisted mode analysis
is due to the ignorance of the localization of NS5-branes.
One may expect that we can obtain
the correct spectrum if we take account of the localization of
NS5-brane and solve the Klein-Gordon equation
on the geometry.
However, we can see that this is not the case as long as
we use the field theory approximation.
The field theory approximation for the tachyon field
seems not to work in this situation and
we probably need to use string theory including excitation modes
in the localized NS5-brane background
in order to obtain the masses correctly.

\section{Conclusions and Discussions}
In this paper, we computed the mass correction to
the twisted modes in deformed orbifolds for several concrete examples.
We found that we can obtain localized tachyons
with masses of order of the curvature turned on
if all the supersymmetries are broken on
the deformed orbifold.

The emergence of tachyonic modes indicates the instability of the spacetime.
Indeed, the tachyonic mode of the deformed ${\bf C}^2/{\bf Z}_2$
obtained in \S\ref{twisted.sec}
corresponds to the instability of the spacetime against the creation
of the Witten's bubble studied in \cite{unstablens5}.
The result in \S\ref{twisted.sec} confirms that
this is only the unstable.

Although we discussed the relation between deformed orbifold
and dual NS5-brane configurations only for
non-supersymmetric ${\bf C}^2/{\bf Z}_{2N_{\rm NS5}(2N_{\rm NS5}+1)}$ case,
it may also be interesting
to study dual descriptions of other unstable orbifold deformations.

The number of localized tachyons
depends on the choice of an orbifold
and the way of deformation.
It may be possible to construct
solitonic objects with these tachyons.
In the ${\bf C}^2/{\bf Z}_2$ case, which we studied in detail,
there is one localized tachyon.
Unfortunately, we cannot use this tachyon field to construct a
domain wall.
The moduli space associated with three blow-up modes of
${\bf C}^2/{\bf Z}_2$ is ${\bf R}^3/{\bf Z}_2$.
Although one of three axes
in ${\bf R}^3$ becomes tachyonic by the deformation making $R_{\rm TN}$ finite,
the positive and the negative directions of the axis are identified
by ${\bf Z}_2$.
Because of this, we have only one direction along which the tachyon condensates
and two different domains cannot be produced.
Thus we cannot use this tachyon to make domain walls.
However, we may be able to obtain localized tachyons available for
the construction of solitonic objects
by different choices of orbifolds and deformations.
The localized tachyons obtained in the ${\bf C}^3/{\bf Z}_3$
and ${\bf C}^4/{\bf Z}_4$ cases are potential candidates.
We hope to come back to the subject elsewhere.

\section*{Acknowledgements}
We thank T.~Kawano,
Y.~Nakayama, Y.~Tachikawa, H.~Takayanagi
for useful discussions. 
Y.~I. is supported in part by
Grant-in-Aid for the Encouragement of Young Scientists
(\#15740140) from the Japan Ministry of Education, Culture, Sports,
Science and Technology,
and by Rikkyo University Special Fund for Research.



\end{document}